\documentclass[prx,preprint,superscriptaddress,doublespace,final,showpacs,showkeys,nobibnotes,titlepage,twoside,10pt]{revtex4-1}
\usepackage{amsmath}
\usepackage{graphicx}
\usepackage{amsfonts}
\usepackage{amssymb}
\usepackage{subfigure}
\usepackage{float}
\usepackage{color}
\usepackage{soul} % Added by TCH 18-March-2016
\begin{document}

\title{Slip avalanches in metallic glasses and granular matter reveal universal dynamics}

\author{D. V. Denisov}
\email{email: d.denisov@uva.nl}
\affiliation{Institute of Physics, University of Amsterdam, P.O. Box 94485, 1090 GL Amsterdam, The Netherlands}
\author{K. A. L\H{o}rincz}
\affiliation{Institute of Physics, University of Amsterdam, P.O. Box 94485, 1090 GL Amsterdam, The Netherlands}
\author{W. J. Wright}
\affiliation{Department of Mechanical Engineering, Bucknell University, One Dent Drive, Lewisburg, PA 17837}
\affiliation{Department of Chemical Engineering, Bucknell University, One Dent Drive, Lewisburg, PA 17837}
\author{T. C. Hufnagel}
\affiliation{Department of Materials Science and Engineering, Johns Hopkins University, Baltimore, MD 21218}
\affiliation{Department of Mechanical Engineering, Johns Hopkins University, Baltimore, MD 21218}
\author{A. Nawano}
\affiliation{Department of Physics, University of Illinois at Urbana Champaign, 1110 West Green Street, Urbana, IL 61801}
\author{X. J. Gu}
\affiliation{Department of Mechanical Engineering, Bucknell University, One Dent Drive, Lewisburg, PA 17837}
\author{J. T. Uhl}
\affiliation{Retired}
\author{K. A. Dahmen}
\affiliation{Department of Physics, University of Illinois at Urbana Champaign, 1110 West Green Street, Urbana, IL 61801}
\author{P. Schall}
\affiliation{Institute of Physics, University of Amsterdam, P.O. Box 94485, 1090 GL Amsterdam, The Netherlands}

\begin{abstract}
Universality in materials deformation is of intense interest: universal scaling relations if exist would bridge the gap from microscopic deformation to macroscopic response in a single material-independent fashion.
While recent agreement of the force statistics of deformed nanopillars, bulk metallic glasses, and granular materials with mean-field predictions supports the idea of universal scaling relations, here for the first time we demonstrate that the universality extends beyond the statistics, and applies to the slip dynamics as well.
By rigorous comparison of two very different systems, bulk metallic glasses and granular materials in terms of both the statistics and dynamics of force fluctuations, we clearly establish a material-independent universal regime of deformation. We experimentally verify the predicted universal scaling function for the time evolution of individual avalanches, and show that both the slip statistics and dynamics are universal, i.e. independent of the scale and details of the material structure and interactions. These results are important for transferring experimental results across scales and material structures in a single theory of deformation.
\end{abstract}

\maketitle

\section{Introduction}
The notion of universality represents a longstanding question in materials deformation, which has been traditionally described by material-specific relations and mechanisms. The existence of universal scaling relations, if confirmed experimentally, would provide a novel means to connect microscopic rearrangements to macroscopic stress-strain response in a single theory of deformation across a wide range of solid materials. Recently, power-law distributions measured in the stress signals of slowly deformed single crystals~\cite{Zaiser2006,Friedman2012}, bulk metallic glasses (BMGs)~\cite{BMGsetupRef,Antonaglia2014}, rocks~\cite{Scholz1968,12Decade}, granular materials~\cite{Granularpaper,Dalton2001,Nori2006,Sumita2009,Godano2011} and even earthquakes \cite{Rice1993,Fisher1997,Wyss2004,Wyss2005} reveal very similar strongly correlated deformation, suggesting underlying universal scaling relations in the slow deformation of solids. These distributions are also well described by a mean-field model of elasto-plastic deformation~\cite{MFpaper}, in which the material's elasticity causes coupling between locally yielding regions resulting in slip avalanches with intermittency as observed in the experiments. A recent comparison of widely different systems showed that indeed the fluctuations of the applied stress follow very similar power-law distributions across a wide range of length scales from nanometers to kilometers~\cite{12Decade}, as adequately described by the mean-field model, lending credence to the idea of an underlying universal mechanism of deformation. Yet, unlike equilibrium critical phenomena, where universality has been rigorously demonstrated by meticulous measurements of scaling relations, such measurements are lacking in the deformation of materials. More importantly, no measurement has yet elucidated the applicability of universal scaling relations
%in the non-equilibrium plastic deformation of materials
to the dynamics of the slip avalanches. Establishing the universality not only of the  statistics, but also of the dynamics of the slip avalanches would provide important grounds for the claim of universality.

In this paper we provide the first rigorous investigation of universal scaling behavior by comparing avalanche {\it statistics} and {\it dynamics} in two very different systems, bulk metallic glasses \cite{BMGpaper} and granular materials \cite{Granularpaper}, in which high resolution stress measurements are possible. We show that despite the large differences in the nature of the two materials in terms of the size, interactions, and dynamics of the constituent particles, they share a regime with identical rescaled stress fluctuations, temporal profiles, and dynamics, all accurately described by mean-field theory. This universal regime results from the system-independent elastic coupling of yielding regions, leading to long-range correlated avalanches of deformation as described by the mean-field model. Besides this universal regime, we also delineate a non-universal regime with system-specific power-law statistics, governed by boundary conditions and finite size effects. This first rigorous comparison of slip statistics and dynamics in two disparate systems corroborates the existence of a universal scaling regime and suggests a universal theory of deformation.

\section{Experimental setup and mean-field model}
Due to their very different nature of hard versus soft solids, metallic glasses and granular materials differ greatly in mechanical properties such as modulus, ductility and elastic strain.  To nevertheless resolve and compare the fine fluctuations of the applied stress in the two systems, we developed specific experimental protocols for each \cite{BMGsetupRef,Granularpaper}.
For the metallic glass specimens we applied uniaxial compression tests using a precisely aligned load train with a fast-response load cell and high-rate data acquisition, see Fig.~\ref{Fig:setup}(a). We used a constant displacement rate with a nominal strain rate of $10^{-4} s^{-1}$, and a bulk metallic glass with composition $Zr_{45}Hf_{12}Nb_{5}Cu_{15.4}Ni_{12.6}Al_{10}$, and specimens 6 mm long along the loading direction with a cross section of 1.5 mm x 2 mm. During compression, the specimen deforms elastically until a shear band or slip event initiates. This causes the displacement rate to temporarily exceed the displacement rate imposed on the specimen, resulting in a stress drop as shown in Fig. \ref{Fig:setup}(c), inset \cite{BMGpaper}. The size of the stress drop is proportional to the slip size. %Subsequently the stress increases until initiation of another slip event. We measure the stress drops in our metallic glass with high temporal resolution to resolve the dynamics of the slip events. This enables us to extract a wide range of predicted scaling exponents and scaling functions that uniquely identify the underlying slip statistics and dynamics.
For the granular system, we used a shear cell with built-in pressure sensors to record the force fluctuations on the tilting walls, as shown in Fig.~\ref{Fig:setup}(b) \cite{Granularpaper}. The granular particles, around $3\cdot10^5$ spheres with a diameter of $d=1.5$ mm and a polydispersity of $\sim5\%$, are confined by a top plate, subjected to confining normal pressure between 4 and 10 kPa, resulting in a particle volume fraction of $55-60\%$. The granular material is sheared at a constant rate $\dot{\gamma}=9.1 \cdot 10^{-4}$ to a total strain of $\gamma=20\%$, and force drops are identified around the monotonically increasing average force, as shown in Fig. \ref{Fig:setup}(d), inset. The number of granular particles is large compared to typical laboratory granular studies, but of course many orders of magnitude smaller than the number of atoms in the metallic glass specimens. The granular linear system size of $\sim70$ particle diameters across can lead to significant truncation of large avalanches and hence to more pronounced finite size effects than for the metallic glass.

To describe the stress fluctuations in both systems, we use a simple analytic model that predicts the slip statistics for elasto-plastic solids~\cite{BMGpaper,MFpaper}. The model assumes that real solids have elastically coupled weak spots, which are known as shear transformation zones in a metallic glass. Each weak spot slips by a random amount when the local stress exceeds a threshold. A slipping weak spot can trigger another weak spot to slip as well in a slip avalanche, causing the intermittent response that is observed in experiments. For bulk metallic glasses and granular materials the model assumes that a recently slipped weak spot is slightly weaker than before, due to dilation~\cite{BMGpaper}. As a result, the model predicts a universal power-law scaling for slip avalanches in a range of sizes that is not affected by finite-size effects of the specimen. Large avalanche slips have different dynamics. They recur almost periodically and span a macroscopic fraction of the system. The model predicts how the average slip avalanche size for the smaller slips grows as a large slip is approached.
From the average slip size it is in principle possible to extract at what stress the next catastrophically large slip will take place. The model also predicts a large number of scaling laws, for example how the statistics changes with applied strain rate and stress, allowing us to extrapolate from one loading condition to another.

%Although in granular experiments the stress-strain curve does not reach a steady state it was shown in \cite{Granularpaper} that this fact does not change the general statistics of the avalanche size distribution. It affects only the distribution tail corresponding to the largest events.

{\begin{figure*}
\centering
~~~~~~~~\includegraphics[width=0.3\textwidth]{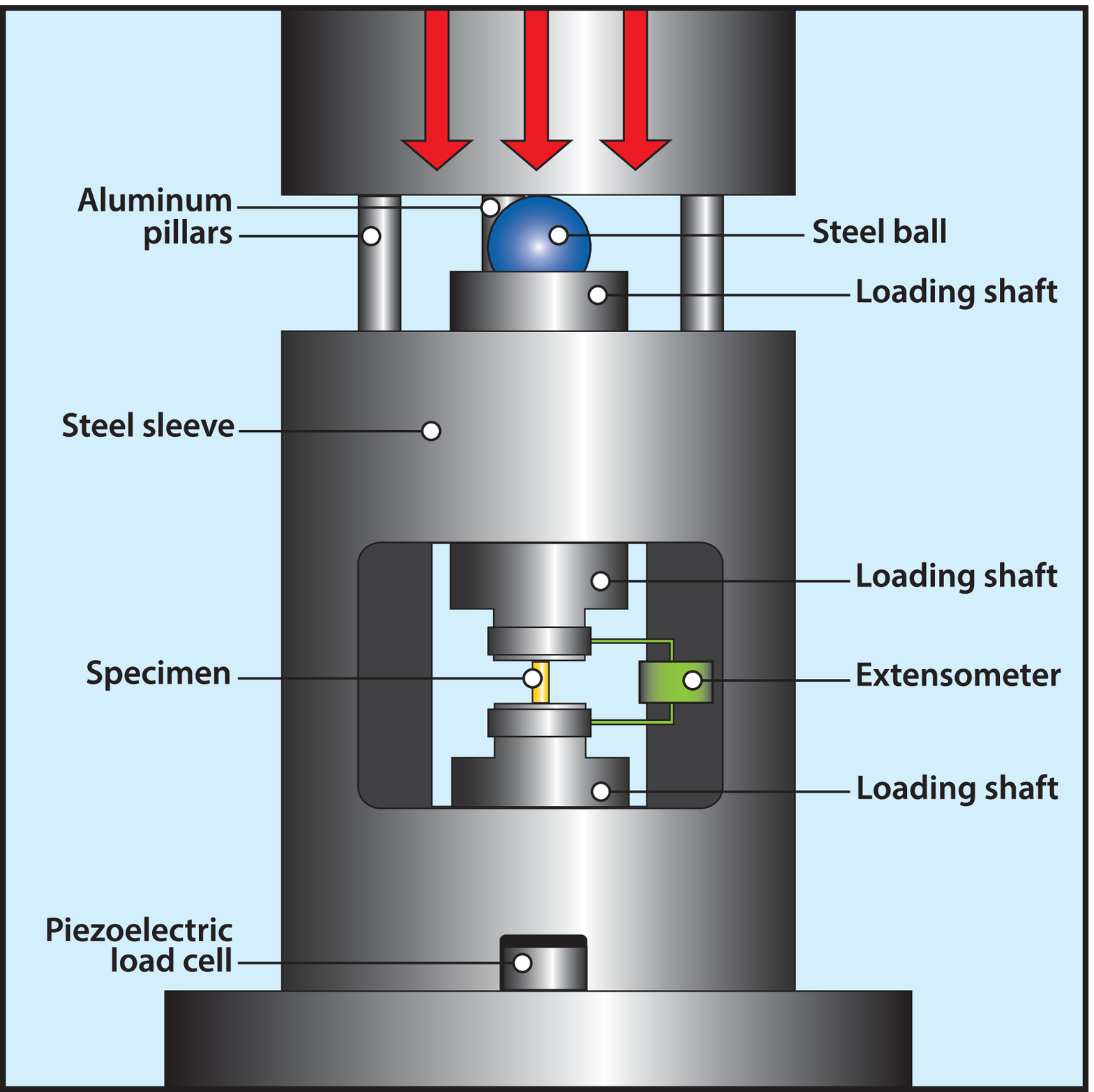}~~~~~~~~~~~
\includegraphics[width=0.375\textwidth]{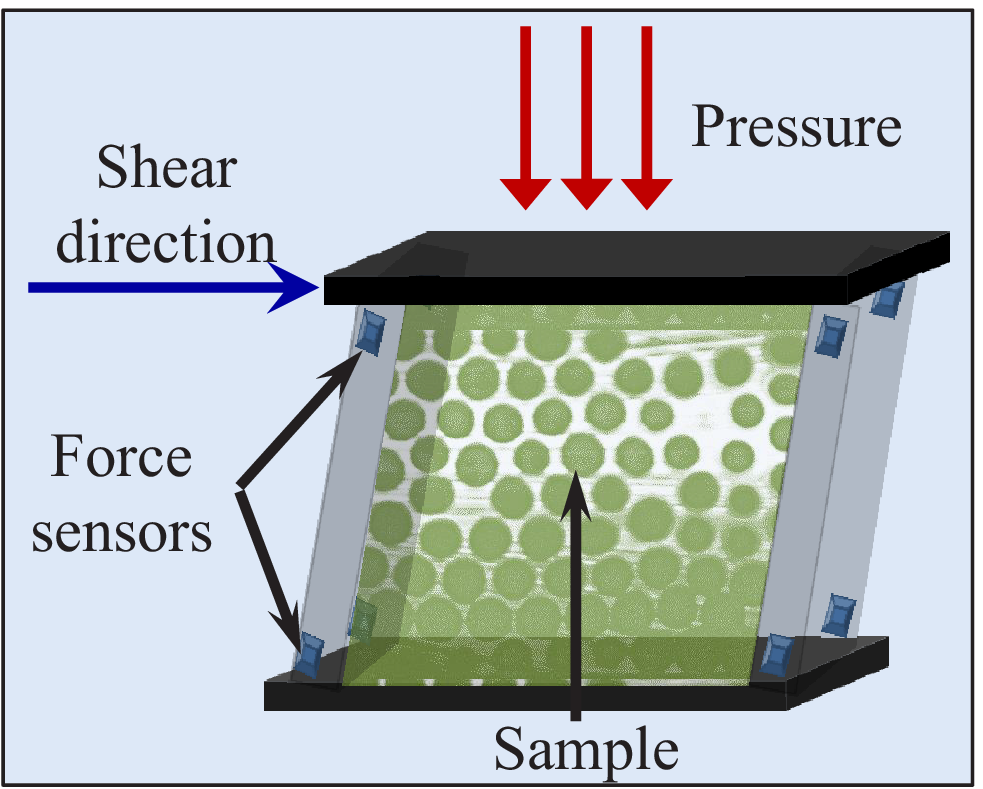}
\includegraphics[width=0.39\textwidth]{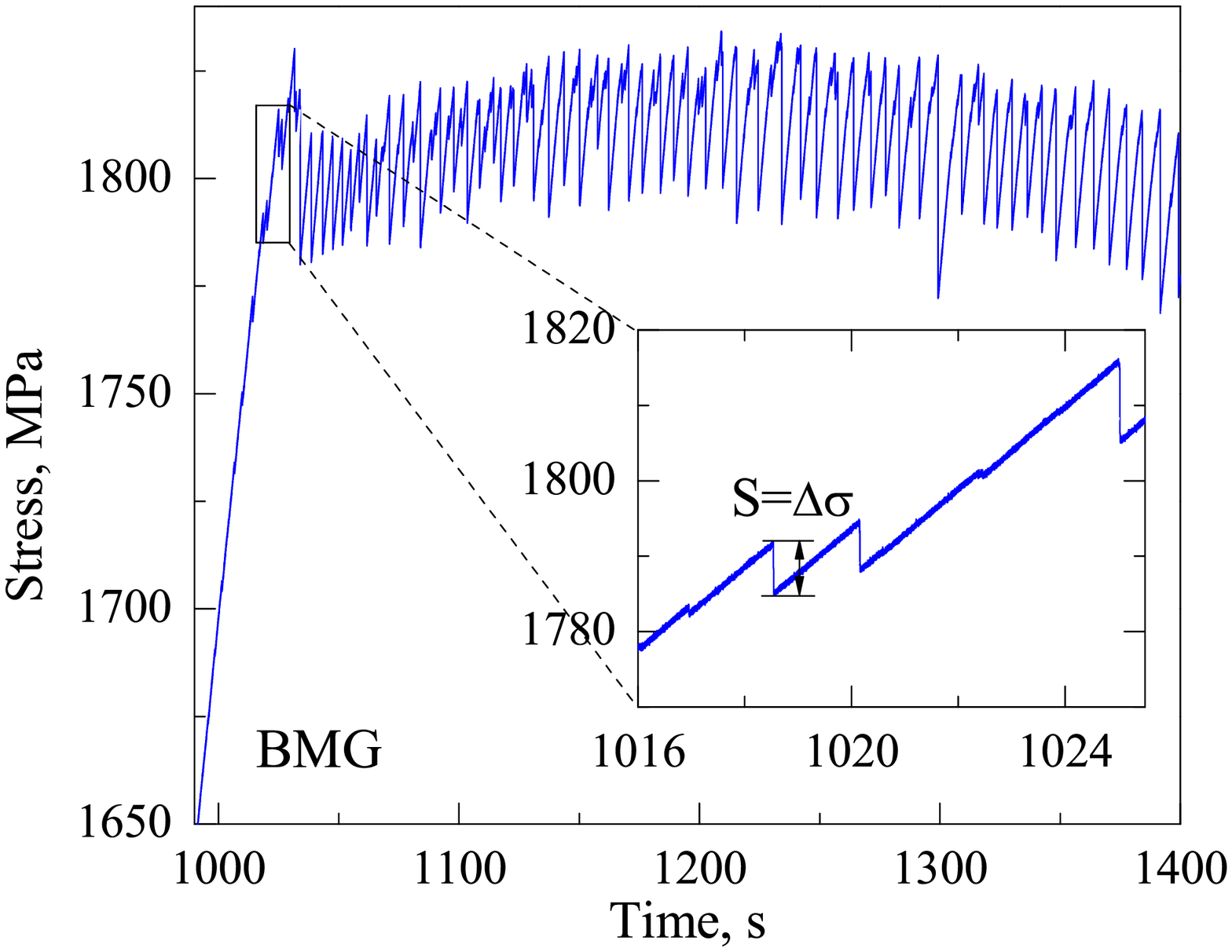}
\includegraphics[width=0.40\textwidth]{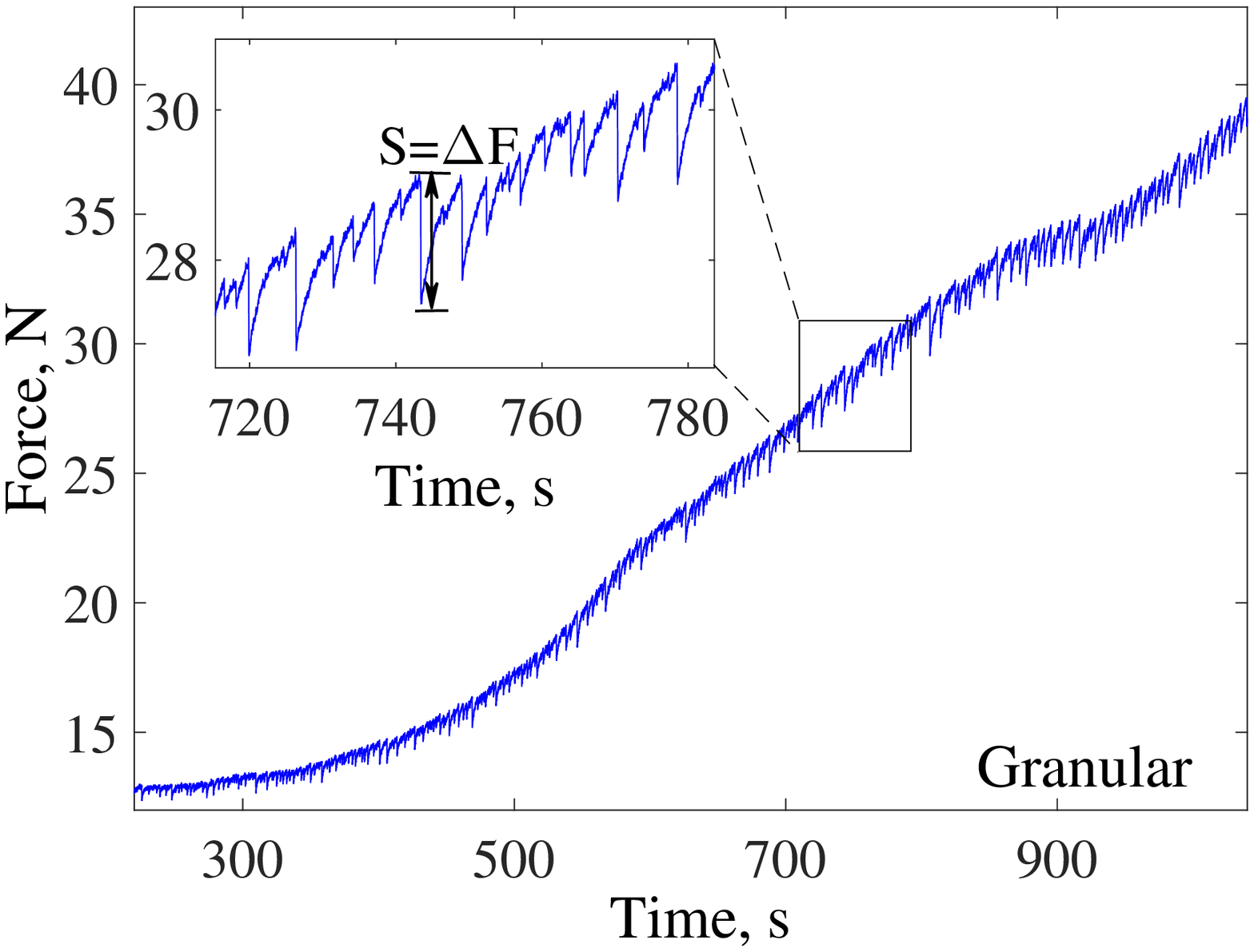}
\begin{picture}(0,0)(0,0)
\put(-405,160){(a)}
\put(-210,160){(b)}
\put(-405,2){(c)}
\put(-210,2){(d)}
\end{picture}
\caption{{\bf Metallic glass and granular setups and measurements.} (a) Schematic of the bulk metallic glass measurement setup. Two tungsten carbide platens that are constrained by a steel sleeve compress the metallic glass specimen. See Ref. \cite{BMGsetupRef} for details. (Drawing is courtesy of Adrienne Beaver, Bucknell University.)
(b) Schematic of the granular shear cell setup with force sensors in the walls. Loads imposed on top exert a constant confining pressure. See Ref. \cite{Granularpaper}.
(c) and (d) Metallic glass and granular data - the main panels show applied stress or force versus time, insets show magnifications of the data. This is the same data as shown in \cite{BMGpaper} and \cite{Granularpaper}.}
\label{Fig:setup}
\end{figure*}

\section{Results}

The raw stress-time data of the two systems in Fig.~\ref{Fig:setup}c and d show pronounced stress fluctuations; rapid stress drops demarcate stress relaxation events, during which the displacement rate temporarily exceeds the displacement rate applied to the specimen.
We define the size of these avalanche events $S$ from the magnitude of sharp stress drops $\Delta\sigma$ and force drops $\Delta F$ in the metallic glass and granular systems, respectively (see insets of Figs. \ref{Fig:setup}c and d).
Due to the very different particle size (angstroms for atoms in the metallic glass opposed to millimeters for the granular particles), and the different nature of interaction (atomic potential versus frictional contacts), the magnitude of stress fluctuations differs greatly, being several ten megapascals for the metallic glass, opposed to several hundred pascals for the granular material.
Nevertheless, we can collapse the stress drop distributions by simple rescaling that accounts for the different stress magnitudes of the hard metallic and soft granular materials. To show this, we plot selected rescaled distributions of stress drop magnitudes and durations in Fig.~\ref{Fig:Cs}.
The probability of stress drops larger than size $S$, known as the complementary cumulative size distribution, is shown in Fig.~\ref{Fig:Cs}a. The metallic glass and granular distributions show excellent overlap for avalanche sizes $S$ in the range $S_{\min}^{GRN}<S<S_{\max}^{GRN}$ that are not affected by the sample boundaries, and exhibit significant deviations for larger avalanches. In the small-avalanche regime, both distributions closely follow a power law $C(S)\sim S^{-(\tau - 1)}$ with exponent $\tau - 1 = 1/2$, in excellent agreement with predictions by the mean field model (dashed line).
%with scaling regime starting at $S_{\min}^{BMG}=0.2MPa$ and $S_{\min}^{GRN}=0.02N$ ($S_{\min}^{BMG}=s_{\min}^{GRN}\cdot10^{-1}\frac{N}{MPa}$).
In particular, the granular data approaches that of the metallic glass and the model predictions with increasing confining pressure that pushes the granulate deeper into the jammed solid regime. This is because, unlike the metallic glass that is held together by attractive molecular interactions, the granular particles are repulsive and held together merely by the applied confining pressure. For larger sizes, the granular power-law distributions are truncated due to finite size effects: they extend over significantly shorter ranges than those of the bulk metallic glass that follows the mean-field prediction up to larger avalanches. In contrast, in the small-avalanche regime not affected by finite size effects, which is the scaling regime of the model, the overlap of the distributions and model predictions is surprisingly good.

In this scaling regime, each weak spot slips only once in an avalanche; these slips are small enough to not be affected by the sample boundaries. In contrast, large avalanches for $S>S_{\max}^{GRN}$ have very different dynamics: they behave similarly to a crack cutting through a macroscopic fraction of the sample, and feel the boundaries of the sample. In the model they are unstoppable or runaway events, where each weak spot slips many times during the same avalanche. The resulting time evolution of slip is very smooth, while for the small avalanches the time evolution of slip is very jerky~\cite{WrightTBP}. Small avalanches are power law distributed because they are always close to stopping, while large avalanches are runaway events that only stop when they have cut through a macroscopic fraction of the system.

The similarity of the avalanche statistics is further confirmed in Fig. \ref{Fig:Cs}b where we show the avalanche duration versus size. Again, for the small avalanches with size $S_{\min}^{GRN}<S<S_{\max}^{GRN}$, the agreement between the metallic glass and granular data is remarkable: both exhibit identical scaling of the size-dependent duration according to $t(S)\sim S^{\sigma\nu z}$ with $\sigma\nu z\simeq1/2$ as predicted by the model. Again, the scaling regime extends to larger avalanches for the metallic glass due to its larger system size. For the granular material, the data crosses over to a second scaling regime $t(S)\sim S^{\tau_{L}}$ with a much smaller exponent $\tau_{L}\simeq0.1$. This much shallower growth of avalanche duration indeed indicates large slip events, common for shear bands or cracks, for which uniform sliding occurs along the entire shear plane. Indeed, it is known that granular materials always shear band~\cite{SB_review}, while metallic glasses exhibit a regime of homogeneous deformation at high temperature and low shear rates.

We further explore the correspondence of avalanches by plotting their duration distribution in Fig. \ref{Fig:Cs}c. We again find good agreement in the small-avalanche regime: the metallic glass and granular systems exhibit identical power-law distributions with the predicted slope of $-1$. Similar to the avalanche size in Fig.~\ref{Fig:Cs}a, the scaling regime extends to large avalanches for the metallic glass, while for the granular system a second scaling regime emerges that clearly changes with the applied confining pressure, and is thus a non-universal regime that depends on the system details.

Another characteristic property of the force drops is the rate of stress release, which gives insight into the propagation dynamics of individual avalanches. Plotting the rate of stress release as a function of avalanche size, we find excellent agreement between the metallic glass and granular data over the entire avalanche regime (Fig. \ref{Fig:Cs}d), signifying that the underlying avalanche propagation dynamics for small avalanches may be the same in both systems.
Yet the scaling range of the stress drop sizes that can be compared to mean field theory is limited by finite size effects.

\begin{figure*}
\centering
\includegraphics[width=0.49\textwidth]{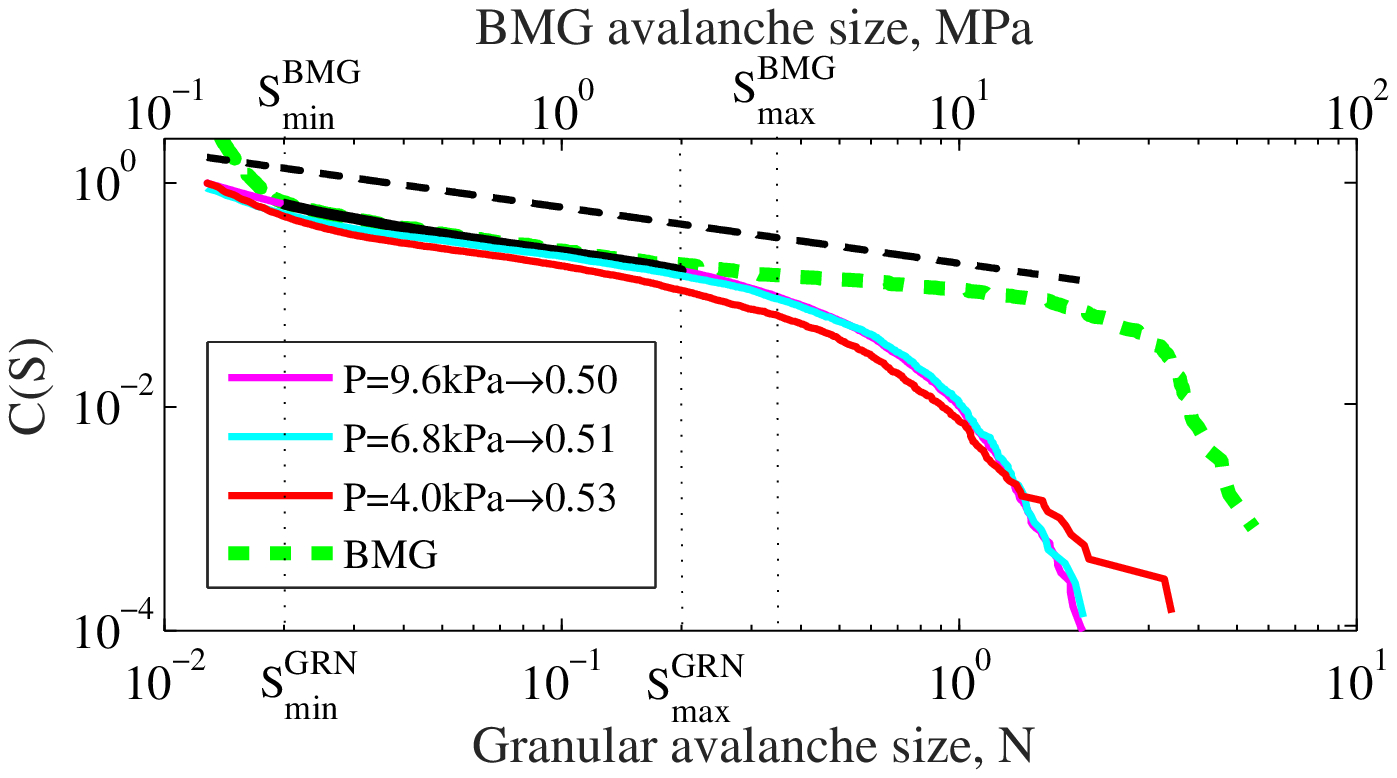}
\includegraphics[width=0.49\textwidth]{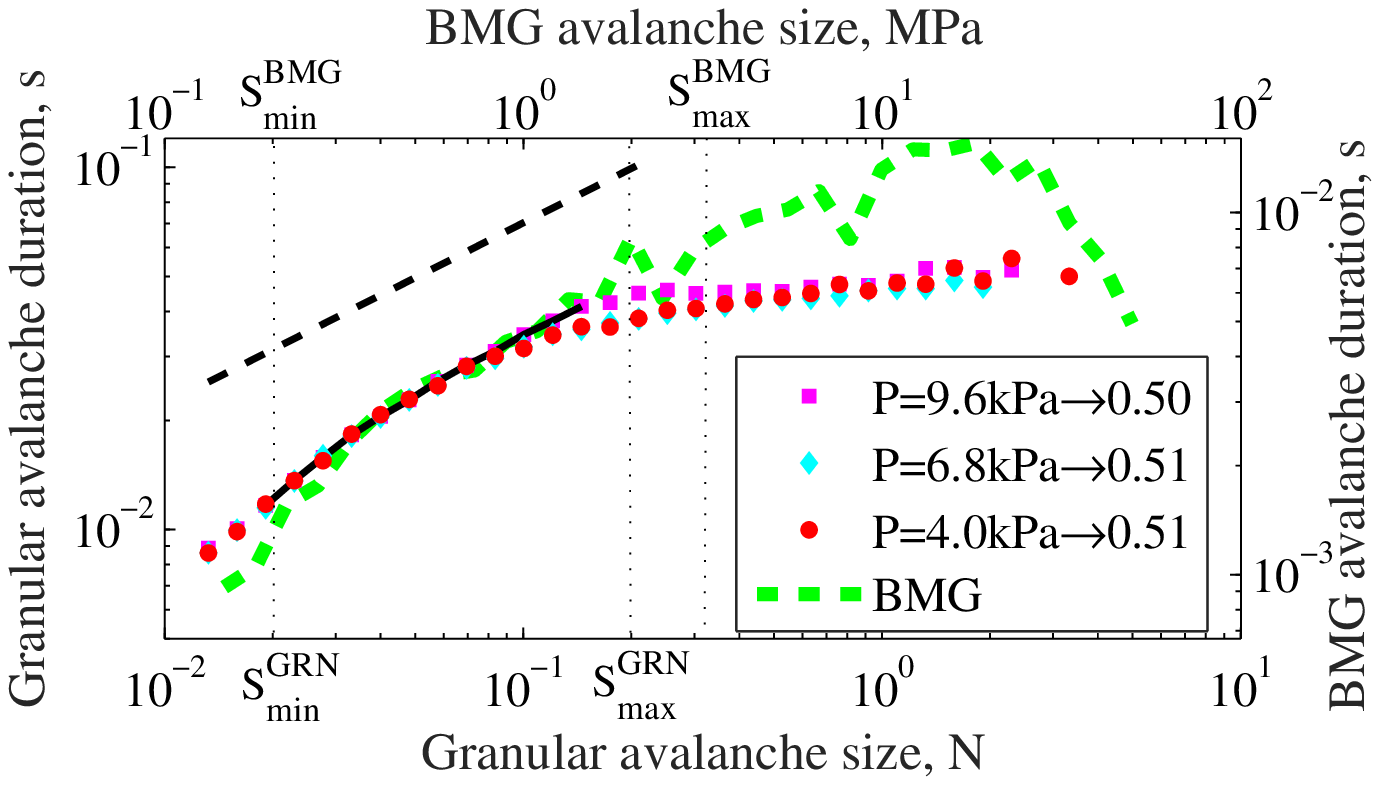} \\
\vspace{0.3cm}
\includegraphics[width=0.49\textwidth]{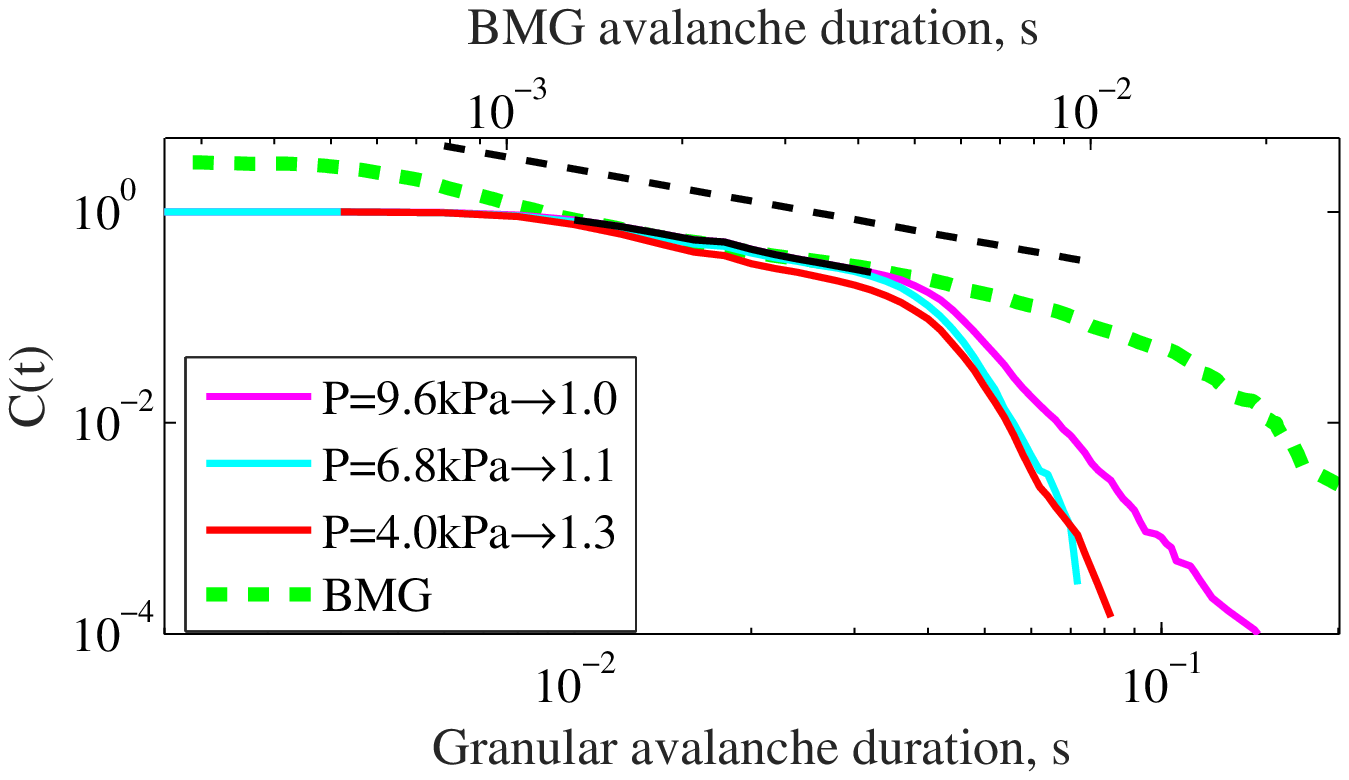}
\includegraphics[width=0.49\textwidth]{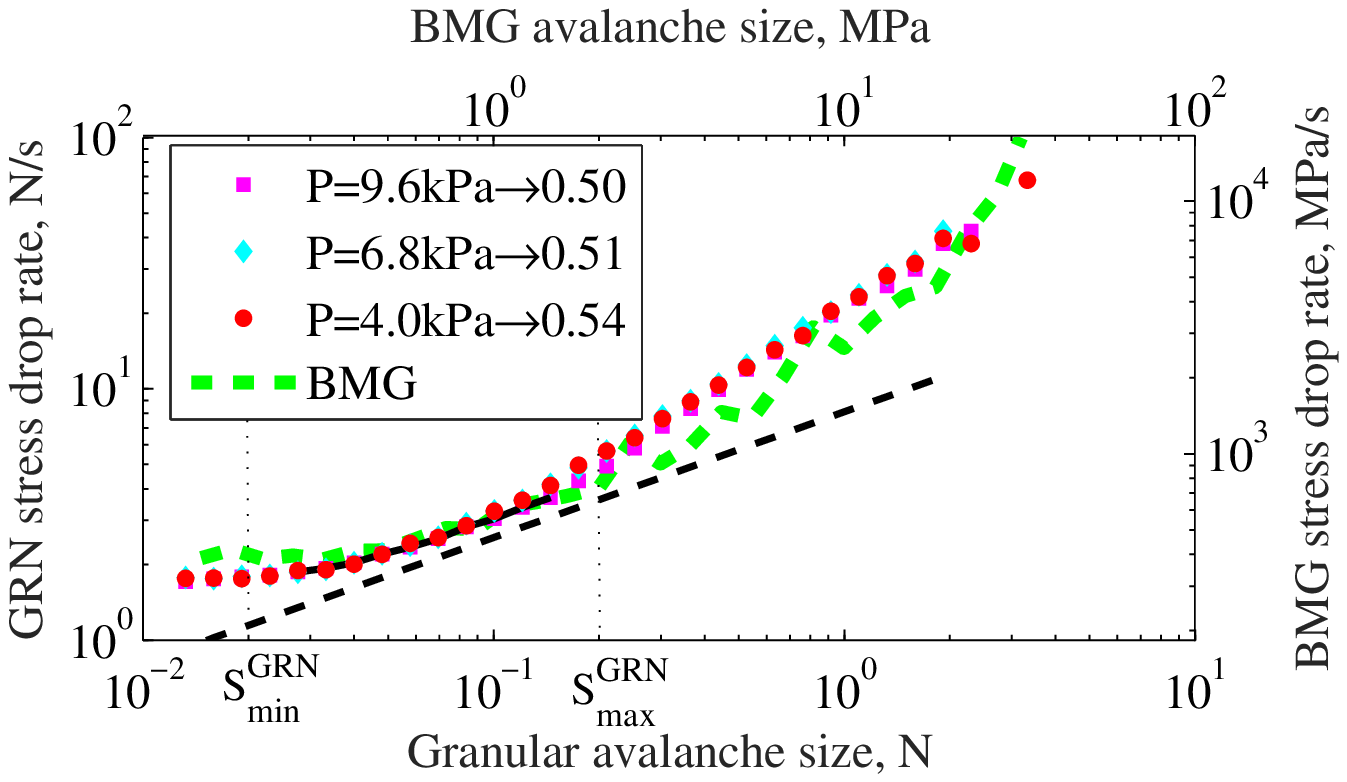}
\begin{picture}(0,0)(0,0)
\put(-495,155){(a)}
\put(-240,155){(b)}
\put(-495,5){(c)}
\put(-240,5){(d)}
\end{picture}
\caption{{\bf Avalanche statistics.} Four scaling parameters are compared for the metallic glass (green dashed line) and the granular material (colored solid lines with color denoting confining pressure):
(a) Complimentary cumulative distribution $C(S)$ of avalanche size,
(b) Avalanche duration versus avalanche size,
(c) Complimentary cumulative distribution $C(t)$ for avalanche duration, and
(d) Stress drop rate versus avalanche size. In each plot the solid black line shows the portion of the granular data corresponding to the scaling regime ($S_{\min}^{GRN}<S<S_{\max}^{GRN}$) for data collected at 9.6~kPa. The dashed black line shows the slope expected from the prediction of the mean field model. The legends show the slope value of the granular curves in the scaling regime for pressures 4.0, 6.8 and 9.6~kPa.}
\label{Fig:Cs}
\end{figure*}

We now take advantage of the finely resolved signals to compare the full time evolution of individual avalanches. We show the rate of force release as a function of time for avalanches in the scaling regime in Fig.~\ref{Fig:Aprof} (for data collected at maximum pressure 9.6 kPa). As expected for the scaling regime, we can indeed collapse all granular avalanche profiles with different durations onto a single master curve as shown in Fig.~\ref{Fig:Aprof}a. A similar collapse for the metallic glass avalanches has been shown in \cite{BMGpaper}. This self-similarity lends credence to the idea that in this regime the system is indeed described by robust scaling relations.

We compare the granular data with that of the metallic glass and mean-field predictions in Fig.~\ref{Fig:Aprof}b. While for the metallic glass, the avalanches show symmetric profiles, in good agreement with mean-field predictions~\cite{BMGpaper}, for the granular material, the avalanche profiles exhibit a slight asymmetry. We associate this asymmetry with delayed damping effects~\cite{Zapperi2005} similar to earthquakes \cite{Mehta2006}. Delay effects can originate with time scales inherent to the friction between the particles. As the granular particles are relatively soft, their elastic relaxation time that sets the microscopic delay time for the onset of slip is considerable. A similar explanation has been suggested for earthquakes \cite{Mehta2006}, and for Barkhausen noise in magnetic materials, where the delay is due to eddy currents in the material \cite{Zapperi2005,Dahmen2005}.
In contrast, in metallic glasses there is no friction between the atoms inside the alloy, and consequently no significant microscopic delay time to yield a noticeable skewing of the avalanche shapes. In either case, the asymmetry of the velocity profile does not affect the scaling exponents; they are still given by the mean field model predictions as shown in \cite{Sethna2001}.

Another difference between the avalanche shapes is that they are not as flattened for the granular material as they are for the metallic glass. The reason for this difference is the limited machine stiffness. For the metallic glass experiments the machine stiffness was chosen to be large~\cite{BMGpaper}, leading to a broadening of the avalanche shape~\cite{BMGpaper,Zapperi2005}. In the granular experiments the walls are of similar stiffness to the granular particles, thus not significantly flattening the temporal profile. By fitting the predicted form with limited machine stiffness to the granular data we find good agreement between model predictions and experiments as shown in Fig.~\ref{Fig:Aprof}b. In summary, our finely resolved measurements of the avalanche profiles reveal the details of machine stiffness and avalanche delay effects due to interparticle friction \cite{Zapperi2005} within the same scaling and universality class.

\begin{figure*}
\centering
\includegraphics[width=0.9\textwidth]{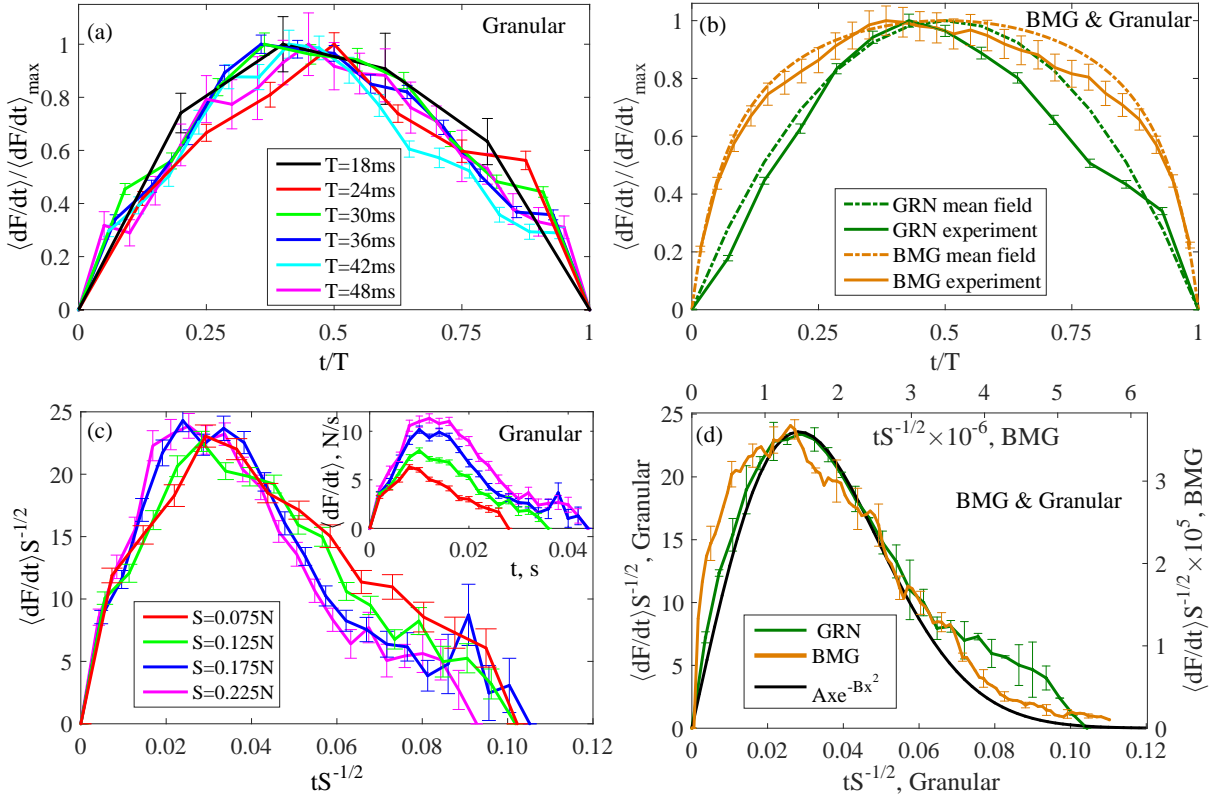}
\caption{
{\bf Avalanche dynamics: Temporal avalanche profiles in the universal scaling regime.}
(a) Average stress-drop rate of the granular system, normalized by the maximum rate. Profiles are averaged over avalanches from small bins of their durations. Error bars are calculated as standard error of the mean.
(b) Stress-drop rates compared for metallic glass (brown) and granular material (green). Solid lines show averages over avalanches in the scaling regime, and dash-dotted curves show mean-field predictions.
(c) Granular stress-drop rates for fixed avalanche sizes $S$ in the scaling regime ($S_{\min}^{GRN}<S<S_{\max}^{GRN}$). Inset shows original data, and main panel shows collapsed profiles scaled along both axes by $S^{-1/2}$.
(d) Comparison of the averaged collapsed profiles for granular data (c) and metallic glass data \cite{BMGpaper}. The black curve shows scaling functions predicted by the mean-field model for both granular materials (bottom axis) and metallic glasses (top axes), which perfectly overlap with each other, further corroborating the similarity of the slip avalanche statistics of metallic glasses and granular materials. The granular fitting constants for the scaling function $Ax\exp(-Bx^2)$ are $A=1.46$ and $B=6.6\times10^{-4}$, the metallic glass constants are $A=3.98\times10^{11}$ and $B=2.18\times10^{11}$.
}
\label{Fig:Aprof}
\end{figure*}

We can further test the scaling relation between avalanche size and duration by collapsing the avalanche profiles as a function of avalanche size. To do so, we sort avalanches according to their size, focusing on the small-avalanche scaling regime. Individual profiles are shown in the inset of Fig. \ref{Fig:Aprof}c. These profiles indeed collapse onto a single master curve when we rescale both axes by $S^{-1/2}$ (main panel of Fig. \ref{Fig:Aprof}c) giving yet another confirmation of the validity of the scaling relation. The average of these profiles also agrees well with that of the metallic glass, and closely matches the prediction by mean-field theory, as shown in Fig. \ref{Fig:Aprof}d. The small difference between metallic glass and granular data for small values of $tS^{-1/2}<0.02$ appears due to differences in particle softness and machine stiffness, similar to Fig. \ref{Fig:Aprof}b. Consequently, the metallic glass profiles can be also fitted with the mean-field theory using slightly different values of the non-universal parameters $A$ and $B$ described in \cite{BMGpaper}, but the form of the two scaling functions $Ax\exp(-Bx^2)$ for granular and metallic glass data can be perfectly overlapped with each other when plotted in their corresponding axes (Fig. \ref{Fig:Aprof}d). While some small deviations for the granular avalanches occur at large values of $tS^{-1/2}>0.06$ due to poor statistics, the agreement of the granular and metallic glass data with the mean field model is striking. We thus conclude that this small-avalanche regime, while limited for the granular system due to finite size effects, has hallmarks of a universal scaling regime.

In contrast, the second scaling regime for large avalanches $S>S_{\max}^{GRN}$ shows non-universal properties that depend on the system geometry and boundary conditions, consistent with the model predictions. Indeed, Figs. \ref{Fig:Cs}a and c already suggest that the granular scaling in this regime depends on the applied confining pressure that drives the system into different jammed states. We can still collapse the avalanche profiles for the granular system using data collected at maximum pressure 9.6 kPa; starting with the raw profiles (inset in Fig.~\ref{Fig:AprofLarge}a), we achieve excellent collapse by scaling the vertical axis by $S^{-1.1}$, as shown in the main panel of Fig. \ref{Fig:AprofLarge}a. Scaling along the horizontal axis is not required due to the almost constant avalanche duration (Fig. \ref{Fig:Cs}b).
Although the predictions that mean field theory makes for the small avalanches are not expected to necessarily extend all the way to the large avalanches that are affected by system boundaries and loading geometries, still the collapsed granular profiles are also in good agreement with the small-avalanche mean-field scaling function $Ax\exp(-Bx^2)$ indicated by the black line.

Interestingly, the details of these large-avalanche profiles allow us to elucidate the connection to the small-avalanche regime: initially ($t<8\cdot10^{-3}$ $s$), these profiles exhibit a "foot" that corresponds precisely to the profile of the small avalanches shown in the inset of Fig. \ref{Fig:Aprof}c. When reaching their maximum $\langle dF/dt \rangle\approx10$ $N/s$ at $t\approx7\cdot10^{-3}$ $s$, some of these small avalanches nucleate into larger ones, upon which the avalanche size grows faster, as clearly seen in the profiles in Fig. \ref{Fig:AprofLarge}a, inset. This nucleation picture of large avalanches is consistent with the mean field model \cite{Fisher1997,MFpaper,Dahmen2009}.

For the metallic glass, this foot is very long~\cite{WrightTBP} (see inset of Fig. \ref{Fig:AprofLarge}b where profiles are centered at the peak positions). Neglecting the foot, we can achieve a reasonably good collapse along the vertical axis by scaling with $S^{-1.6}$, which is quite far from the mean-field scaling $S^{-0.5}$ for the small avalanches. This is not surprising, since the mean-field theory predicts that the $S^{-0.5}$ scaling only applies to the small avalanches but not for the large ones.
We associate this difference in the granular and the metallic glass behavior with the difference in boundary effects and loading conditions of the two systems: because the large avalanches feel the system boundaries, and the boundary conditions in both experiments are different, it is expected that the large avalanche profiles in these systems look different. We hence identify this empirical scaling behavior in the large-avalanche regime as non-universal and system-specific, and even changing with confining pressure for the granular system. The nucleation of large avalanches from the small ones and its dependence on the internal and external conditions is an interesting topic for further studies.

\begin{figure*}
\centering
\includegraphics[width=0.9\textwidth]{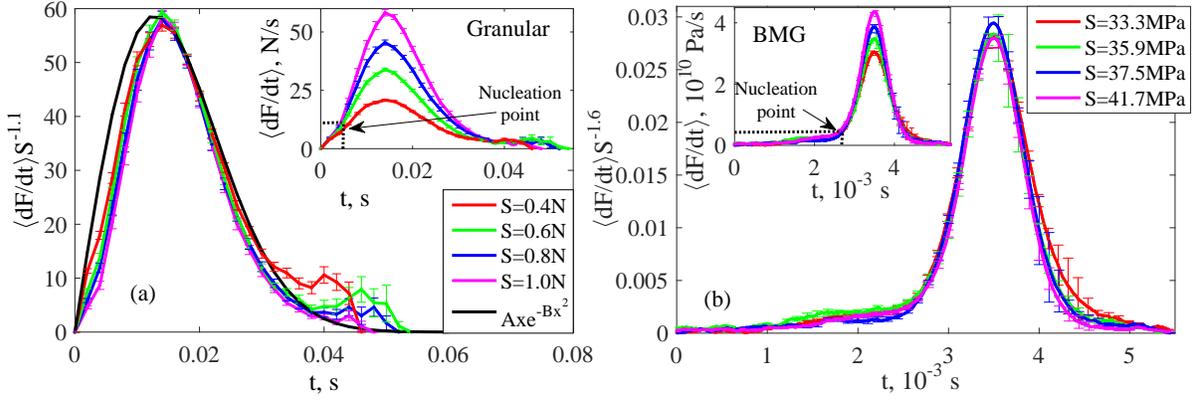}
\caption{
{\bf Temporal profiles of large granular avalanches.} Stress-drop rate profiles of large avalanches for the granular (a) and metallic glass system (b). Profiles averaged by size are shown in the insets, and collapsed data is shown in the main panels. Error bars are calculated as standard error of the mean.
(a) Large granular avalanches show good collapse when scaling the vertical axis by $S^{-1.1}$. Scaling along the horizontal axis is not required. The mean-field scaling function $Ax\exp(-Bx^2)$ can be fitted with $A=7.4$ and $B=3\times10^{-3}$. The nucleation point at $t\approx7\cdot10^{-3}$ $s$ and $\langle dF/dt \rangle\approx10$ $N/s$ when the small avalanche turns into a large one is very close to a maximum point of the small avalanche profiles shown in Fig. \ref{Fig:Aprof}c.
(b) Large metallic-glass avalanches show collapse when the vertical axis is scaled by $S^{-1.6}$. Along the horizontal axis, the profiles have been centered manually at the peak positions. In the large-avalanche regime, the scaling of the avalanches is not universal, as it is affected by the different boundary conditions of the systems, in agreement with mean-field predictions.
}
\label{Fig:AprofLarge}
\end{figure*}

\section{Conclusion}
We have demonstrated universal features of slip avalanches in metallic glasses and granular systems. Due to the very different particle size and interaction of the hard atomic and soft granular amorphous solids, the stress fluctuations are orders of magnitude different. Yet, their distributions reveal strikingly similar statistics and dynamics with identical power-law behavior of avalanche sizes and durations. For the metallic glass this scaling regime spans a relatively broad range; for the granular material, besides the universal regime, we also observe a non-universal scaling regime, characterized by large system-spanning avalanches that appear to have almost identical duration. We attribute the large avalanches to shear bands or crack-like slip. Most importantly, for the scaling regime, we observe clear universal behavior in the scaling exponents and avalanche dynamics of the small avalanches that do not span across the entire sample. The detected slight differences in the avalanche profiles arise from delayed damping effects in the granular materials due to friction, which differs from the particle interactions in bulk metallic glasses.
%In granular systems, the grains interact via friction, while in metallic glasses the interactions are interatomic interactions, which are of very different nature. Both can, however, be accounted for by the mean field model with different amount of delay effects, making
In this way the large, asymmetric granular avalanche profiles are similar to those of large-scale earthquakes.

These results provide an important step towards a universal understanding of the deformation of amorphous materials. While previous studies showed that slowly-compressed single crystals, bulk metallic glasses, rocks, granular materials, and the earth all deform via intermittent slips or "quakes" \cite{12Decade}, the current study for the first time compares not only scaling exponents but also scaling functions for the dynamics of slip across two systems with completely different scales, structures, and interactions. The good agreement between the systems, and between the experimental data and mean-field predictions for avalanche statistics {\it and} dynamics significantly expands the claim that these systems may be described by a unifying theory not only with respect to their slip statistics, but also with respect to the slip dynamics.

\bibliographystyle{prsty}

\section{References}

\begin{thebibliography}{15}
\bibitem{Zaiser2006} M. Zaiser, \textit{Scale invariance in plastic flow of crystalline solids}, Adv. Phys. \textbf{55}, 185–245 (2006).
\bibitem{Friedman2012} N. Friedman, \textit{et al}, \textit{Statistics of dislocation slip-avalanches in nanosized single crystals show tuned critical behavior predicted by a simple mean field model}, Phys. Rev. Lett. \textbf{109}, 095507 (2012).
\bibitem{BMGsetupRef} W. J. Wright, M. W. Samale, T. C. Hufnagel, M. M. LeBlanc, and J. N. Florando, \textit{Studies of shear band velocity using spatially and temporally resolved measurements of strain during quasistatic compression of a bulk metallic glass}, Acta Mater. \textbf{57}, 4639 (2009).
\bibitem{Antonaglia2014} J. Antonaglia. \textit{et al}, \textit{Tuned Critical Avalanche Scaling in Bulk Metallic Glasses}, Sci. Rep. \textbf{4}, 4382/1–5 (2014).
\bibitem{Scholz1968} C. H. Scholz, \textit{The Frequency-magnitude relation of microfracturing in rock and its relation to earthquakes}, Bull. Seismol. Soc. Am. \textbf{58}, 399–415 (1968).
\bibitem{12Decade} J. T. Uhl, S. Pathak, D. Schorlemmer, X. Liu, R. Swindeman, B. A. W. Brinkman, M. LeBlanc, G. Tsekenis, N. Friedman, R. Behringer, D. Denisov, P. Schall, X. J. Gu, W. J. Wright, T. Hufnagel, A. Jennings, J. R. Greer, P. K. Liaw, T. Becker, G. Dresen, and K. A. Dahmen, \textit{Universal Quake Statistics: From Compressed Nanocrystals to Earthquakes}, Sci. Rep. \textbf{5}, 16493 (2015).
\bibitem{Granularpaper} D. V. Denisov, K. A. Lorincz, J. T. Uhl, K. A. Dahmen, and P. Schall, \textit{Universality of slip avalanches in flowing granular matter}, Nat. Comm. \textbf{7}, 10641 (2016).
\bibitem{Dalton2001} F. Dalton, and D. Corcoran, \textit{Self-organized criticality in a sheared granular stick-slip system}, Phys. Rev. E \textbf{63}, 061312 (2001).
\bibitem{Nori2006} M. Bretz, R. Zaretzki, S. B. Field, N. Mitarai, and F. Nori, \textit{Broad distribution of stick-slip events in Slowly Sheared Granular Media: Table-top production of a Gutenberg-Richter-like distribution}, Europhys. Lett. \textbf{74}, 1116 (2006).
\bibitem{Sumita2009} N. Higashi, I. Sumita, \textit{Experiments on granular rheology: Effects of particle size and fluid viscosity}, J. Geophys. Res. \textbf{114}, B04413 (2009).
\bibitem{Godano2011} M. P. Ciamarra, E. Lippiello, L. de Arcangelis, and C. Godano, \textit{Statistics of slipping event sizes in granular seismic fault models}, EPL \textbf{95}, 54002 (2011).
\bibitem{Rice1993} Y. Ben-Zion, and J. R. Rice, \textit{Slip patterns and earthquake populations along different classes of faults in elastic solids}, J. Geophys. Res. \textbf{98}, 14109–14131 (1993).
\bibitem{Fisher1997} D. Fisher, K. A. Dahmen, S. Ramanathan, and Y. Ben-Zion, \textit{Statistics of earthquakes in simple models of heterogeneous faults}, Phys. Rev. Lett. \textbf{78}, 4885–4888 (1997).
\bibitem{Wyss2004} D. Schorlemmer, S. Wiemer, and M. Wyss, \textit{Earthquake statistics at Parkfield: 1. Stationarity of b-values}, J. Geophys. Res. \textbf{109}, B12307/1–17, doi: 10.1029/2004JB003234 (2004).
\bibitem{Wyss2005} D. Schorlemmer, S. Wiemer, and M. Wyss, \textit{Variations in earthquake-size distribution across different stress regimes}, Nature \textbf{437}, 539–542 (2005).
\bibitem{MFpaper} K. A. Dahmen,  Y. Ben-Zion, and J. T. Uhl, \textit{A simple analytic theory for the statistics of avalanches in sheared granular materials}, Nat. Phys. \textbf{7}, 554 (2011).
\bibitem{BMGpaper} J. Antonaglia, W. J. Wright, X. J. Gu, R. R. Byer, T. C. Hufnagel, M. LeBlanc, J. T. Uhl, and K. A. Dahmen, \textit{Bulk Metallic Glasses Deform Via Slip Avalanches}, Phys. Rev. Lett. \textbf{112}, 155501 (2014).
\bibitem{WrightTBP} W. J. Wright, Y. Liu, X. J. Gu, K. D. Van Ness, S. L. Robare, X. Liu, J. Antonaglia, M. LeBlanc, J. T. Uhl, T. C. Hufnagel, and K. A. Dahmen, \textit{Experimental evidence for both progressive and simultaneous shear during quasistatic compression of a bulk metallic glass}, J. App. Phys. \textbf{119}, 084908 (2016).
\bibitem{SB_review} P. Schall and M. van Hecke, \textit{Shear Bands in Matter with Granularity}, Ann. Rev. Fluid Mech. {\bf 42}, 67 (2010).
\bibitem{Zapperi2005} S. Zapperi, C. Castellano, F. Colaiori, and G. Durin, \textit{Signature of effective mass in crackling-noise asymmetry}, Nat. Phys. \textbf{1}, 46 (2005).
\bibitem{Mehta2006} A. P. Mehta, K. A. Dahmen, and Y. Ben-Zion, \textit{Universal mean moment rate profiles of earthquake ruptures}, Phys. Rev. E \textbf{73}, 056104 (2006).
\bibitem{Dahmen2005} K. A. Dahmen, \textit{Nonlinear dynamics: Universal clues in noisy skews}, Nat. Phys. \textbf{1}, 13 (2005).
\bibitem{Sethna2001} J. P. Sethna, K. A. Dahmen, and C. R. Myers, \textit{review article Crackling noise}, Nature \textbf{410}, 242 (2001).
\bibitem{Dahmen2009} K. A. Dahmen, Y. Ben-Zion, and J. T. Uhl, \textit{Micromechanical model for deformation in solids with universal predictions for stress strain curves and slip avalances}, Phys. Rev. Lett. \textbf{102}, 175501 (2009).

%\bibitem{Mehta2002} A. P. Mehta, A. C. Mills, K. A. Dahmen, and J. P. Sethna, Phys. Rev. E \textbf{65}, 046139 (2002).
\end{thebibliography}

\section{Acknowledgements}
This work is part of the research program of FOM (Stichting voor Fundamenteel Onderzoek der Materie), which is financially supported by NWO (Nederlandse Organisatie voor Wetenschappelijk Onderzoek); NSF DMR 1042734 (WJW); NSF DMS 1069224, NSF DMR 1005209  and NSF CBET 1336634 (KD). TCH acknowledges support from the National Science Foundation under grant DMR 1408686. We also thank the Kavli Institute for Theoretical Physics and the Aspen Center of Physics for hospitality and support via grants NSF PHY 1125915  and NSF PHY 1066293 respectively.

\section{Author contribution}
D.V.D. and P.S. designed the granular research, D.V.D. performed the granular measurements, D.V.D. and K.A.L. analyzed the granular data. W.J.W. designed the metallic glass experiments, X.J.G. performed the metallic glass measurements, A.N. analyzed the metallic glass data, D.V.D compared the data for the metallic glass and granular systems. J.T.U. and K.A.D. derived the theoretical predictions and guided the metallic glass data analysis and comparison to the model predictions. P.S. and D.V.D wrote major parts of the manuscript, with contributions from W.J.W, K.D, and T.C.H.

\section{Competing financial interests}
The authors declare no competing financial interests.

\end{document}